\documentclass[reqno]{article}
\usepackage{graphicx}
\usepackage{amsmath}

\input{tcilatex}
\bigskip

\begin{document}

\title{Some Algebro-Geometric Aspects of The SL(2, R) Wess-Zumino-Witten Model of
Strings on an ADS$_{3}$ Background}
\author{Bogdan G. Dimitrov \\
{\small \textit{Bogoliubov Laboratory for Theoretical Physics}}\\
[-1.mm] {\small \textit{Joint Institute for Nuclear Research, Dubna 141 980,
Russia}}\\
[-1,mm] {\small \textit{email: bogdan@thsun1.jinr.ru}}}
\date{}
\maketitle

\begin{abstract}
The SL(2, R)\ WZW model of strings on an ADS3 background is investigated in
the spirit of \ J.Maldacena's and H.Ooguri's approach (hep-th/0001053\ and
hep-th/0005183). Choosing a standard, but most general three-variable
parametrization of the SL(2, R) group element g, the system of equations for
the Operator Product Expansion (OPE) relations is analysed. In the
investigated SL(2, R)\ case, this system is consistent if each three points
on the complex plane lie on a certain hypersurface in CP3. A system of three
nonlinear first-order differential equations has been obtained for the
parametrization functions. It was demonstrated also how the mathematical
apparatus of generalized functions and integral geometry can be implemented
in order to modify the integral operators, entering the Kac-Moody and
Virasoro algebras.

\end{abstract}

\section{Introduction}

\bigskip Several years ago serious attempts were made to apply the
originally developed in [1, 2] two-dimensional conformal WZW model to the
model of strings and branes on a curved background. The basic idea was to
''supply'' the group element $g$, entering the WZW action, with the
properties of the chosen background Anti De-Sitter spacetime ADS$_{3}$ by
means of a suitable parametrization of $g$ in terms of the ADS$_{3}$ global
coordinates [3, 4, 5]. However, the original WZW\ model, defined on an
(unspecified) two-dimensional world-sheet spacetime, turned out to be not
well suited to account for some specific effects, such as stretchings and
windings of strings \ close to the boundary of ADS$_{3}.$ In search of a
resolution, two main approaches were proposed. The\textbf{\ first one} was
based on the notion of a \textbf{spectral flow}, which generates new
solutions of the equations of motion for SL(2, R) by acting on standard
representations by elements of the loop group. This amounts to performing a
prescribed transformation of the global ADS coordinates (for example, $t$
and $\Phi $, $\Phi $- periodic)\thinspace in respect to the string
world-sheet coordinates ($\tau $,$\sigma $) [3, 5]. The \textbf{second
approach}, developed in [6], had the purpose to ''redefine'' the integral
formulation of the conserved charges $J_{0}^{a}=\oint dzJ^{a}(z)$, where $%
J^{a}(z)$ are the conformal currents, entering the OPE\ relations.The
important idea was to multiply the integrand $J^{a}(z)$ by a \textbf{%
holomorphic vertex operator} $\gamma ^{n}(z)$, so that $J_{n}^{a}=\oint
dzJ^{a}(z)\gamma ^{n}(z)$. In such a way, the zero-mode symmetry of the
algebra $\left[ J_{0}^{a},J_{0}^{b}\right] =if_{c}^{ab}J_{0}^{c}$ was
extended to the infinite symmetry of the affine Lie algebra in spacetime%
\textbf{\ } $\left[ L_{m},J_{n}^{a}\right] =-nJ_{n+m}^{a}$ and $\left[
J_{n}^{a},J_{m}^{b}\right] =if_{c}^{ab}J_{n+m}^{c}+\frac{k}{2}n\delta
^{ab}\delta _{n+m,0}$. \textbf{The peculiarity of this approach was that the
Kac-Moody and Virasoro algebras should not be considered as automatically
following from the OPE relations}, but their consistent resolution is needed
[6].

The purpose of the present paper is to introduce another modification of the
integral conformal currents, based on integrating on a contour not on the
two-dimensional world-sheet, but on the contour, obtained as an intersection
of a set of hyperplanes (inside the ADS hyperboloid) with the hyperboloid.
This is the integral geometric approach of Gel'fand, Graev and Vilenkin [7],
which recently was applied also in [8] to construct a non-local map from
de-Sitter space to Anti de-Sitter, which commutes with the $SO(d,1)$
isometry group. The objective of the present paper also is to explore the
OPE\ relations for thye SL(2, R) case and written in terms of the three
SL(2, R) parametrization variables. More details about the present approach
are to be found in [10].

\section{A Three-Parameter SL(2,R) Parametrization of the WZW Model in ADS$%
_{3}$ Background}

Our starting point is the WZW effective action:

\begin{equation}
S=\frac{1}{4\lambda ^{2}}\int Tr(\partial _{\mu }g^{-1}\partial _{\mu
}g)d^{2}\xi +k\Gamma (g)\text{ ,}  \tag{2.1}
\end{equation}
where $g$ is the group element, $\lambda $ and $k$ are dimensionless
coupling constants, $\xi =(\xi _{1},\xi _{2})$ are the coordinates of the
two-dimensional world-sheet and $\Gamma (g)$ is the boundary WZW term:
\begin{equation}
\Gamma (g)=\frac{1}{24}\int d^{3}X\text{ }\epsilon ^{\alpha \beta \gamma
}Tr(g^{-1}\partial _{\alpha }gg^{-1}\partial _{\beta }gg^{-1}\partial
_{\gamma }g)\text{ \ .}  \tag{2.2}
\end{equation}
The integration is performed over a three-dimensional ball with coordinates $%
X^{\alpha }$, the boundary of which is identified with the two-dimensional
world-sheet. Since the third de-Rham cohomology vanishes for the SL(2, R)
group, in the present case it shall be neglected.

In terms of the global coordinates $(X^{0},X^{1},X^{2},X^{3})$ on the Anti
de-Sitter hyperboloid
\begin{equation}
-(X^{0})^{2}-(X^{3})^{2}+(X^{1})^{2}+(X^{2})^{2}=-L^{2}\text{ ,}  \tag{2.3}
\end{equation}
where $L$ is the de - Sitter radius, one can parametrize the SL(2, R) group
element $g$ as
\begin{equation}
g=\frac{1}{L}\left(
\begin{array}{cc}
X^{0}+X^{1} & X^{2}+X^{3} \\
X^{2}-X^{3} & X^{0}-X^{1}
\end{array}
\right)  \tag{2.4}
\end{equation}
so that $detg=1$.

Let us now introduce three angle variables $(\Phi $,$\Psi ,\Theta )$, which
shall parametrize the SL(2,R) model. According to a well -known theorem [9],
each SL(2,R) group matrice $g$ can be represented as
\begin{equation}
g=d_{1}(-e)^{\varepsilon _{1}}s^{\epsilon _{2}}pd_{2}\text{ \ ,}  \tag{2.6}
\end{equation}
where $\epsilon _{1},\epsilon _{2}=0$ or $1$, $d_{1}$, $d_{2}$, $(-e)$ and $%
s $ are the diagonal matrices
\begin{equation}
d_{1}=\left(
\begin{array}{cc}
e^{-\Phi } & 0 \\
0 & e^{\Phi }
\end{array}
\right) \text{ }\ \text{;\ \ \ \ \ }d_{2}=\left(
\begin{array}{cc}
e^{-\Psi } & 0 \\
0 & e^{\Psi }
\end{array}
\right) \text{ \ \ ; \ \ \ }(-e)=\left(
\begin{array}{cc}
-1 & 0 \\
0 & -1
\end{array}
\right) \text{ }\ \text{;\ }s=\text{\ \ \ }\left(
\begin{array}{cc}
0 & 1 \\
-1 & 0
\end{array}
\right)  \tag{2.7}
\end{equation}
The matrix $p$ in the case represents the rotations in the hyperbolic plane:
\begin{equation}
p=\left(
\begin{array}{cc}
\cosh \Theta & \sinh \Theta \\
\sinh \Theta & \cosh \Theta
\end{array}
\right)  \tag{2.8}
\end{equation}
and $-\infty <\Theta <+\infty $ . Depending on the values of $\epsilon _{1}$
and $\epsilon _{2}$, four different cases have to be considered. Since for
both the cases ($\varepsilon _{1}=1,\varepsilon _{2}=0$ - Case I) and for ($%
\varepsilon _{1}=\varepsilon _{2}=0-$ Case II) one can write
\begin{equation}
(-e)^{\varepsilon _{1}}s^{\varepsilon _{2}}=\varepsilon E\text{ \ \ \ ,}
\tag{2.9}
\end{equation}
where $E$ is the unit matrix and $\varepsilon =\pm 1$, one can easily find
the elements of the SL(2, R) matrix
\begin{equation}
g=d_{1}(-e)^{\varepsilon _{1}}s^{\varepsilon _{2}}pd_{2}=\varepsilon \left(
\begin{array}{cc}
e^{-\Phi -\Psi }cosh\Theta & e^{-\Phi +\Psi }sinh\Theta \\
e^{\Phi -\Psi }sinh\Theta & e^{\Phi +\Psi }cosh\Theta
\end{array}
\right) \text{ \ \ .}  \tag{2.10}
\end{equation}
Similarly, for the other two cases ($\varepsilon _{1}=\epsilon _{2}=1-$Case
III) and ($\varepsilon _{1}=0$, $\varepsilon _{2}=1-$ Case IV) one obtains
correspondingly the matrices for $(-e)^{\epsilon _{1}}s^{\varepsilon _{2}}$
\ and \ $(-e)^{\epsilon _{1}}s^{\varepsilon _{2}}$ , but these cases shall
not be considered .

Comparing the elements of the matrices (2.5) and (2.10), the global
coordinates on the de Sitter hyperboloid can be expressed as follows:
\begin{equation}
X^{0,1}=\varepsilon \frac{L}{2}cosh\Theta (e^{-\Phi -\Psi }\pm e^{\Phi +\Psi
})\text{ \ \ \ ;\ \ \ \ \ \ \ }X^{2,3}=\varepsilon \frac{L}{2}sinh\Theta
(e^{-\Phi +\Psi }\pm e^{\Phi -\Psi })\text{\ \ }  \tag{2.11}
\end{equation}
and the $+$ sign is for the $X^{0}$ and $X^{2}$ coordinates and the $-$ sign
- for $X^{1}$ and $X^{3}$.

\section{Gauge Currents of the WZW\ Model in Terms \ of \ the $\ \ $SL(2,R)
Parametrization Variables}

\bigskip\ \ \ Let us first define complex coordinates $z$ and $\overline{z}$
on the two-dimensional world-sheet by setting up

\begin{equation}
z=\xi _{1}+i\xi _{2}\ \ \ \ \ \ \ \ \ \ \ \ \ \ \ \ \ \ \overline{z}=\xi
_{1}-i\xi _{2}  \tag{3.1}
\end{equation}
and thus the functional in (2.1) is defined over some two-simensional
Riemann surface.

\bigskip\ \ \ \ The WZW model has a chiral $SL(2,R)\times SL(2,R)$ symmetry,
characterized by an infinite number of conserved currents $J$ and $\overline{%
J}$, derivable from the equations
\begin{equation}
\partial _{\overline{z}}J=0\text{ \ \ \ \ \ \ \ }\partial _{z}J=0  \tag{3.2}
\end{equation}
\ \ \ \ \ \ \ \ $\ \ \ \ \ \ \ \ \ \ $The left and right conformal currents
\ $J_{L}^{a}$ and $J_{R}^{a}$ can be expressed as
\begin{equation}
J_{L}^{a}=kTr(T^{a}g^{-1}\partial _{\overline{z}}g)\text{ \ \ \ \ \ \ \ \ \
\ \ \ \ \ \ \ \ \ \ \ }J_{R}^{a}=kTr(T^{a}\partial _{z}g.g^{-1})  \tag{3.3}
\end{equation}
and $T^{a}$ are the generators of the $SL(2,R)$ algebra, expressible through
the Pauli matrices $\sigma _{1},\sigma _{2},\sigma _{3}$ as $T^{3}=-\frac{i}{%
2}\sigma _{2}=\left(
\begin{array}{cc}
0 & -\frac{1}{2} \\
\frac{1}{2} & 0
\end{array}
\right) $, $T^{+}=\frac{1}{2}(\sigma _{3}+i\sigma _{1})=\left(
\begin{array}{cc}
\frac{1}{2} & \frac{i}{2} \\
\frac{i}{2} & -\frac{1}{2}
\end{array}
\right) $ and $T^{-}=\frac{1}{2}(\sigma _{3}-i\sigma _{1})=\left(
\begin{array}{cc}
\frac{1}{2} & -\frac{i}{2} \\
-\frac{i}{2} & -\frac{1}{2}
\end{array}
\right) $.

Now an important point should be clarified. In the initial works on CFT
[1,2], there was no need to specify the two - dimensional subspace. But
later on, when the WZW model of ADS$_{3}$ strings came out, it appeared
natural, in the framework of some approximation, to relate this
two-dimensional subspace to a two-dimensional subspace of the ADS$_{3}$
spacetime, expressed in global coordinates as
\begin{equation}
ds^{2}=-cosh^{2}\rho dt^{2}+d\rho ^{2}+sinh^{2}\rho d\varphi ^{2}\text{ ,}
\tag{3.4}
\end{equation}
where $\varphi $ is a $2\pi -$ periodic coordinate. In such an approach, $z$
and $\overline{z}$ would be defined as $z=t+i\varphi ,$\ $\overline{z}%
=t-i\varphi $ or as in [5], $t$ and $\varphi $ may be considered to be the%
\textit{\ left- }and \textit{right- }moving coordinates on the world-sheet,
i.e. $t=u+v$ and\ $\varphi =u-v$.\ In all the above case and in many others,
frequently used in the literature, the prescribed dependence between the
parametrization variables and the global ones represents an approximate
approach, chozen for convenience.

Since the global and the parametrization variables $(\Phi ,\Psi ,\Theta )$
(assumed to be different from the ADS global coordinates $%
(X^{0},X^{1},X^{2},X^{3})$ or $(t,\rho ,\varphi )$) depend on the string
world-sheet coordinates, one may write
\begin{equation}
X^{\mu }\equiv X^{\mu }(t,\rho ,\varphi )\equiv X^{\mu }(\Phi (z,\overline{z}%
),\Psi (z,\overline{z}),\Theta (z,\overline{z}))  \tag{3.5}
\end{equation}

where $\mu =0,1,2,3.$ The change of the global coordinates $X^{\mu }\equiv
X^{\mu }(t,\rho ,\varphi )$ from one system to another may be investigated
on a purely algebraic bases, provided that the ADS hyperboloid equation is
given in terms of $X^{\mu }$. As it may be shown [10], if one has the metric
in the form \ (3.4), there is a whole class of functions $X^{\mu }(t,\rho
,\varphi ),$ which satisfy the hyperboloid equation and which can be found
as solutions of a complicated system of \textit{nonlinear differential
equations in partial derivatives. }This is the \textit{first basic aspect}
in the \textbf{algebraic approach}, which has not been developed until now,
because usually for convenience it is preferred to choose a definite
parametrization of the global coordinates $X^{\mu }.$ There is also a
\textit{second aspect} of the algebraic approach, the mathematical
foundation of which is developed in [11]. Suppose one works in different ADS
global coordinates, including also the case when the lenght interval is
expressed through the two-dimensional world-sheet coordinates. Then, if the
metric (3.4) is given, the differentials of all possible global ADS
coordinates have to satisfy a \textbf{cubic algebraic equation}, which
effectively determines all possible parametrizations of the ADS space-time
in terms of the given coordinates $(t,\rho ,\varphi \dot{)}$. As a partial
case [11], the \textbf{elliptic Weierstrass function} and its derivative can
parametrize (satisfy) the parametrizable form of the cubic equation which is
obtained from the original one after applying a linear-fractional
transformation. However, the full ''classification'' of \ all possible
parametrizations on the base of the cubic algebraic equation represents a
complicated and still unresolved mathematical problem.

Concerning the dependence of the global coordinates $X^{\mu }$ on the
parametrization variables $X^{\mu }\equiv X^{\mu }(\Phi (z,\overline{z}%
),\Psi (z,\overline{z}),\Theta (z,\overline{z}))$, evidently some relation
or equations are needed. Further this problem shall be resolved by assuming
that the \textbf{gauge WZW\ currents }are at the same time also \textbf{%
conformal currents}, therefore satisfying the Operator Product Expansion
(OPE)\ relations, but excluding the OPE\ relations with the energy-momentum
tensor. The above mentioned assumption is in fact frequently used in the
literature, although it is not commented at all. In principle, the two types
of currents are different because the \textbf{gauge currents} reflect the
gauge symmetries of the WZW action, while the \textbf{conformal currents}
and more especially the OPE\ relations follow from the \textbf{conformal
Ward identities [}1, 2\textbf{], }which characterize the admissable
transformations of the complex variable under the action of the conformal
group. The identification of \ the \textbf{gauge }with the \textbf{conformal
}currents relates the global ADS space-time symmetries with the string
world-sheet symmetries, which in a sense is similar to the identifications
of the global ADS coordinates with the world-sheet ones. It may be supposed
that in a \textbf{self-consistent }WZW model of strings on an ADS3
background, the parametrization variables \textbf{should not }in advance be
identified with the the global ADS coordinates (or with any combination of
theirs), but rather should be found as a result of an combination of \ both
the \textbf{conformal theory approach} and the \textbf{algebraic theory
approach}. This, however, is very difficult, so as a first step these
approaches shall be presented separately.

Let us now use expressions (2.11) and (3.3) and the formulae for the inverse
matrix $g^{-1}$ (under the condition of $SL(2,R)$ parametrization $detg=1$),
found from (2.4), in order to write down the corresponding \textit{WZW gauge
currents} $J_{R}^{a}$ for $a=3,+,-$ :
\begin{equation}
J_{R}^{3}=a_{1}\frac{\partial \Theta }{\partial z}+a_{2}\frac{\partial \Psi
}{\partial z}\text{ \ ; \ }J_{R}^{+}=b_{1}\frac{\partial \Phi }{\partial z}%
+b_{2}\frac{\partial \Psi }{\partial z}+b_{3}\frac{\partial \Theta }{%
\partial z}\text{ ; \ }J_{R}^{-}=b_{1}\frac{\partial \Phi }{\partial z}+%
\overline{b}_{2}\frac{\partial \Psi }{\partial z}-b_{3}\frac{\partial \Theta
}{\partial z}  \tag{3.6}
\end{equation}
where $a_{1},a_{2},b_{1},b_{2},b_{3}$ are the expressions
\begin{equation}
a_{1}\equiv -ksinh2\Phi \text{ \ ;\ \ \ }a_{2}\equiv ksinh2\Theta cosh2\Phi
\text{ \ ;}  \tag{3.7}
\end{equation}
\begin{equation}
\text{ \ }b_{1}\equiv -L^{2}\text{ ;\ \ }b_{2}\equiv -L^{2}[cosh2\Theta
+isinh2\Theta sinh2\Phi ]\text{\ \ ;\ }b_{3}\equiv iL^{2}cosh2\Phi \text{\ \
.}  \tag{3.8}
\end{equation}
\textbf{\ }

\bigskip Note the appearence of an imaginary part in the currents $%
J_{R}^{\mp }$, while the currents $J_{R}^{3}$ contain only a real part.

\section{An\ Algebraic\ Relation\ From\ The\ Three-Point\ Operator\ Product\
Expansion\ For\ The\ SL(2,R)\ Case}

Let the currents $J_{R}^{3},J_{R}^{+}$ and $J_{R}^{-}$ satisfy \ the OPE
relations for a conformal theory with an affine $SL(2,R)\times SL(2,R)$ Lie
algebra symmetry at level $k$. Since the OPE relations with the energy -
momentum tensor contain derivatives, they shall not be used in the present
investigation, but taking them into account might be an interesting problem
for further research.

In their general form, the OPE relations with the conformal current (only)
can be written as
\begin{equation}
J^{A}(z)J^{B}(w)=\frac{1}{2}\frac{k\eta ^{AB}}{(z-w)^{2}}+\frac{i\epsilon
^{ABC}\eta _{CD}}{z-w}J^{D}\text{,}  \tag{4.1}
\end{equation}
where the indices $A,B,C,D=3,+,-$, $\varepsilon _{ABC\text{ }}$ are the
structure costants of the $SL(2,R)\;$group, $\eta ^{AB}$ is the $SL(2,R)$
group metric - in the case $\eta ^{AB}=diag.(+1,+1,-1)$ and only $\epsilon
_{012}=1$ (the rest ones are zero). For values of $(A,B)=(3,+),(3,-)$ and $%
(+,-)$, the corresponding relations are
\begin{equation}
J^{3}(z)J^{+}(w)=-\frac{iJ^{-}(w)}{z-w}\text{ \ ; \ \ \ \ \ }%
J^{3}(z)J^{-}(w)=-\frac{iJ^{+}(w)}{z-w}\text{ \ ; \ \ \ \ \ }%
J^{+}(z)J^{-}(w)=\frac{iJ^{3}(w)}{z-w}\text{\ \ .}  \tag{4.2}
\end{equation}

For the case with equal indices, the OPE relations are
\begin{equation}
J^{3}(z)J^{3}(w)=\frac{k}{2(z-w)^{2}}\text{ \ \ ; \ \ }J^{+}(z)J^{+}(w)=%
\frac{k}{2(z-w)^{2}}\text{ \ \ ;\ \ \ }J^{-}(z)J^{-}(w)=-\frac{k}{2(z-w)^{2}}%
\text{\ \ \ .}  \tag{4.3}
\end{equation}

Multiplying both sides of the first equality in (4.2) by $J^{-}(v)$ and
making use of \ the third equality in (4.3) (but for the points $w$ and $v$%
), the right hand side (R.H.S.) may be rewritten as
\begin{equation}
J^{3}(z)J^{+}(w)J^{-}(v)=-\frac{iJ^{-}(w)J^{-}(v)}{z-w}=\frac{ik}{%
2(z-w)(w-v)^{2}}  \tag{4.4}
\end{equation}
Again, using (4.2) with the purpose to rewrite $J^{+}(w)J^{-}(v)$ in the
left hand side (L.H.S) of (4.4) and afterwards comparing both sides of
(4.4), a simple algebraic relation is obtained. In the same manner, the
second equality in (4.2) can be multiplied by $J^{+}(v)$ and after
transformation of the L.H.S., another algebraic relation is obtained. The
two simple algebraic relations are the following
\begin{equation}
\frac{1}{(z-v)^{2}}=\frac{1}{(z-w)}\frac{1}{(w-v)}\text{ \ ; \ \ }\frac{1}{%
(w-v)^{2}}=-\frac{1}{(z-w)}\text{\ \ \ .}  \tag{4.5}
\end{equation}
Next, the third relation in (4.2) can be multiplied by $J^{3}(v),$ but the
obtained algebraic relation will be the same as the second relation in
(4.5). Therefore, making use of the above two relations (4.5), one final
algebraic relation will be obtained
\begin{equation}
\frac{1}{(z-v)^{2}}+\frac{1}{(w-v)^{2}}=0\text{ \ \ ,}  \tag{4.6}
\end{equation}
representing a three-dimensional algebraic surface in the complex projective
plane CP3.

\section{Conformal OPE\ \ Relations,\ WZW\ Currents And A System of
Nonlinear Differential Equations For The\ SL(2,R) Parametrization \ Variables%
}

The already found \textbf{WZW currents} $J^{3}(z)$, $J^{+}(z)$, $J^{-}(z)$
(3.6) depend on the symmetries of the WZW action and on the chosen $SL(2,R)$
parametrization variables. As explained in Section 3, it shall be assumed
that \textbf{the WZW\ currents are also conformal currents, satisfying the
OPE\ relations, written for the }$SL(2,R)$\textbf{\ group}. The difficulty
with the OPE\ relations is that they are functional relations, taken at
\textbf{different} world-sheet points.This is the reason these relations do
not give much information, they are discussed usually in regard to the
\textbf{Kac-Moody} and \textbf{Virasoro} algebras, which naturally follow
from them and which shall also be used further. A basic new moment in this
paper will be to use the algebraic relation (4.6), which relates different
points on the world sheet and thus might allow to obtain from the OPE\
relations some equations, taken at one and the same point. In order to
achieve this, some additional assumptions have to be made.

The basic assumption for a \textbf{conformal field theory is that it is
invariant under Mobius transformations on the complex plane}, and the Mobius
transformation is
\begin{equation}
v(z)=\frac{az+b}{cz+d}\text{ \ \ \ \ \ \ \ }a,b,c,d\subset C\text{ \ \ \ \ }%
ad-bc=1\text{ \ .}  \tag{5.1}
\end{equation}

The assumption about Mobius invariance is consistent with the latest
developments in conformal field theory [12,13], since Mobius invariance
turns out to be a more fundamental concept than the conformal structure of
the theory itself. Mobius invariance is enough to define the amplitudes, the
vertex operators and the Virasoro algebra in a conformal field theory . In
the present case, the transformation (4.6) can be written as $z=i\varepsilon
w+(1-i\varepsilon )v$, and if $v$ and $w$ can interchange their places, from
the new relation and the previous it is obtained that $z=pw$ ($p$ - a
complex number)$.$ Therefore, our transformation turns out to be a partial
case of the Mobius one. However, note that the interchanging of places
should be treated as an approximation and not as a natural operation - for
example, from (4.1) it can be obtained by substracting that $\left[
J^{A}(z),J^{B}(w)\right] =2\frac{i\epsilon ^{ABC}\eta _{CD}}{z-w}J^{D}\neq 0$
in the general case.

Now it shall naturally be assumed that the global ADS coordinates $X^{\mu }$
are invariant in respect to the Mobius transformation $w(z)$, i.e.
\begin{equation}
X^{\mu }(z)\equiv X^{\mu }(w(z))\text{ \ and }\ \ \frac{\partial X^{\mu }(z)%
}{\partial z}\equiv \frac{\partial X^{\mu }(w(z))}{\partial z}\ \ \text{\ ,}
\tag{5.2}
\end{equation}
which of course is reminiscent of string world-sheet invariance. Indeed, if
one substitutes the group element $g$ in the WZW\ action, the resulting
action will be for a nonlinear sigma model, where the global ADS coordinates
will represent the string coordinates, for which the reparametrization
invariance requirement is valid. Also, the Mobius invariance concept does
not contradict the existence of the algebraic relation (4.6). The important
moment is that it can be generalized for $n+1$ points, meaning that the
coefficients $a_{n+1}$, $b_{n+1}$, $c_{n+1}$, $d_{n+1}$ in the Mobius
transformation for the $(n+1)$ point will depend on the coefficients of the
remaining $n$ points. However, the number of these $n$ points (with freely
varying coefficients) may go to infinity, so Mobius invariance may hold for
all points. If \ (5.2)\ holds and if the derivatives $\frac{\partial X^{\mu
}(z)}{\partial \Phi }$ ,$\frac{\partial X^{\mu }(z)}{\partial \Psi }$ $\ $%
and $\frac{\partial X^{\mu }(z)}{\partial \Theta }$ \ are \textbf{arbitrary}%
, then it can easily be proved [10] that it is natural to choose \textbf{the
parametrization variables }$(\Phi ,\Psi ,\Theta )$\textbf{\ to be Mobius and
reparametrization invariant}, i.e.
\begin{equation}
(\Phi (w(z)),\Psi (w(z)),\Theta (w(z)))\equiv (\Phi (z),\Psi (z),\Theta (z))%
\text{ \ }\Rightarrow \text{\ }\frac{\partial \Phi (w(z,v))}{\partial w}=%
\frac{\partial \Phi (w(z,v)}{\partial z}\frac{\partial z}{\partial w}%
=-i\varepsilon \frac{\partial \Phi (z)}{\partial z}\text{.}  \tag{5.4}
\end{equation}
In (5.4) $\frac{\partial z}{\partial w}$ can be found from the algebraic
relation (4.6). The same of course holds for the derivatives of $\ \Psi $
and $\Theta $. Note that the assumption about \textbf{arbitrariness} of the
derivatives is substantial - in such an \textbf{approximation} the \textbf{%
conformal} and the\textbf{\ algebraic} approach should be considered as
\textbf{independent one from another}, otherwise instead of (5.5) another
more complicated expression should be used [10]. By means of (4.6), the WZW
currents from formulaes (3.8), taken at the point $w,$ can be expressed
though the currents at the point $z$ in the following simple way:
\begin{equation}
J_{R}^{3}(w)=-i\varepsilon J_{R}^{3}(z)\text{ \ ;}\ \ \ \ \ \ \ \text{\ }%
J_{R}^{+}(w)=-i\varepsilon J_{R}^{+}(z)\text{ ;\ \ \ \ }\ \ \ \text{\ }%
J_{R}^{-}(w)=-i\varepsilon J_{R}^{-}(z)\text{\ \ \ .}  \tag{5.5}
\end{equation}
Substituting these conformal currents into the OPE relations (4.2) for $%
J^{3}(z)J^{+}(w)$ and $J^{3}(z)J^{-}(w)$, it can be derived \
\begin{equation}
J^{+}(z)=\varepsilon _{1}J^{-}(z)\text{ \ \ \ \ \ \ \ \ \ \ \ \ \ \ \ \ }%
J^{3}(z)=i\varepsilon _{2}J^{+}(z)\text{ \ \ \ \ ,}  \tag{5.6}
\end{equation}
where $\varepsilon _{1}$ and $\varepsilon _{2}$ can take values $\pm 1$
independently one from another. If the above equations are written in terms
of the derivatives $\frac{\partial \Phi }{\partial z}$, $\frac{\partial \Psi
}{\partial z}$ and $\frac{\partial \Theta }{\partial z}$, two of the
derivatives $\frac{\partial \Phi }{\partial z}$ and $\frac{\partial \Psi }{%
\partial z}$ can be expressed through the third one $\frac{\partial \Theta }{%
\partial z}$
\begin{equation}
\frac{\partial \Phi }{\partial z}=Q\frac{\partial \Theta }{\partial z}\text{
\ \ \ \ \ \ \ \ \ \ \ \ \ \ \ \ \ \ \ \ \ \ \ \ \ \ }\frac{\partial \Psi }{%
\partial z}=P\frac{\partial \Theta }{\partial z}\text{ \ \ \ \ \ \ ,}
\tag{5.7}
\end{equation}
where $Q$ and $P$ depend on the expressions $b_{1},b_{2}$ and $b_{3}$ and
therefore on the parametrization variables. Next, if one uses the other set
of \ (the first two) OPE equations (4.3) for $J^{3}(z)J^{3}(w)$ and $%
J^{+}(z)J^{+}(w)$, two more relations for $\frac{\partial \Theta }{\partial z%
}\frac{\partial \Psi }{\partial z}$ and $\frac{\partial \Phi }{\partial z}%
\frac{\partial \Psi }{\partial z}$ can be obtained \ and also, a third \
equation for $J^{-}(z)J^{-}(w).$ Combining the obtained system of \textbf{%
three (quadratic) algebraic relations }in respect to the derivatives $\
\frac{\partial \Phi }{\partial z}$, $\frac{\partial \Psi }{\partial z}$ and $%
\frac{\partial \Theta }{\partial z}$ with (5.7), finally one receives the
following \textbf{system of first-order nonlinear differential equations}
for $\Theta $, $\Phi $ and $\Psi $ \bigskip :
\begin{equation}
\frac{\partial \Theta }{\partial z}=T(\Theta ,\Phi ,\Psi )\text{ ; \ \ \ \ }%
\frac{\partial \Phi }{\partial z}=Q(\Theta ,\Phi ,\Psi )T(\Theta ,\Phi ,\Psi
)\text{\ \ \ \ \ \ \ ; }\frac{\partial \Psi }{\partial z}=P(\Theta ,\Phi
,\Psi )T(\Theta ,\Phi ,\Psi )\text{ \ \ .}  \tag{5.8}
\end{equation}
where \bigskip $T(\Theta ,\Phi ,\Psi )$ is a definite function.

\section{ KaC-Moody And\ Virasoro \ Algebras\ And\ Application of the
Integral \ Geometry Approach Of\ Gel'fand, Graev\ And\ Vilenkin}

The definition an expression of the conformal currents $J_{R}^{3}(\zeta )$, $%
J_{R}^{+}(\zeta )$ and $J_{R}^{-}(\zeta )$ and the knowledge of the
conformal stress -energy tensor $T(\zeta )$ is the first step towards
constructing the relevant \textbf{integral quantities}, which in the
notations of \ [2] can be written as
\begin{equation}
L_{n}A_{j}(z,\overline{z})=\oint\limits_{C}T(\zeta )(\zeta -z)^{n+1}A_{j}(z,%
\overline{z})d\zeta \text{ \ ; \ \ }J_{n}^{a}A_{j}(z,\overline{z}%
)=\oint\limits_{C}J^{a}(\zeta )(\zeta -z)^{n}t_{j}^{a}A_{j}(z,\overline{z}%
)d\zeta  \tag{6.1}
\end{equation}
where $n$ can admit positive and negative values, $A_{j}(z,\overline{z})$ is
some field, depending on the holomorphic and anti-holomorphic variables and $%
t_{j}^{a}$ are the group generators. In this standard CFT formulation the
integration is performed on an arbitrary contour $C$ on the complex plane.

\textbf{It is important to note that in standard \ CFT, provided that the
conformal currents }$J^{a}$\textbf{\ satisfy the OPE\ relations, the
integral operators }$L_{n}$\textbf{\ and }$J_{n}^{a}$\textbf{\ will satisfy
the Virasoro algebra }
\begin{equation}
\left[ L_{n},L_{m}\right] =(n-m)L_{n+m}+\frac{1}{12}c(n^{3}-n)\delta _{n+m}%
\text{ \ ; \ \ \ \ \ \ \ }\left[ L_{n},J_{m}^{a}\right] =-mJ_{n+m}^{a}\text{%
\ \ }  \tag{6.2}
\end{equation}
and \textbf{the Kac-Moody algebra }
\begin{equation}
\left[ J_{n}^{a},J_{m}^{b}\right] =f^{abc}J_{n+m}^{c}+\frac{1}{2}kn\delta
^{ab}\delta _{n+m,0}\text{ \ \ .}  \tag{6.3}
\end{equation}

\textbf{Previously we have related the CFT formulation and the WZW
formulation of string theory on an ADS}$_{3}$ \textbf{background} \textbf{\
by assuming that the WZW currents represent the conformal currents as well.
Therefore it is natural to ask whether it is possible to relate the two
formulations also on the level of the integral representation in terms of
the Virasoro and Kac-Moody algebras. }For this purpose,\textbf{\ } it shall
be proposed to introduce an additional contour of integration, obtained as a
result of the intersection of a set of hyperplanes with the ADS hyperboloid,
and this is one of the realizations of the ADS\ (Lobachevsky) space. The
basic problem, which shall be treated further is whether under the
additionally imposed integration the Virasoro and the Kac-Moody algebras
will still hold. \textbf{The mathematical formulation of the additional
contour integration will be given in terms of generalized functions, defined
on some hypersurface [7], and the geometrical meaning might be also
essentially related to the notion of integration on the orisphere (of the
Lobachevsky space), developed in the framework of integral geometry [7] by
Gel'fand, Graev and Vilenkin.}

Let us introduce an \textbf{additional contour integration} $C_{1}$as the
intersection of the set of hyperplanes with the ADS hyperboloid in the
expressions (6.1) for the integral operators $L_{n}$ and $J_{n}^{a}$. In the
present case, since there will be an additional dependence on $\eta =(\eta
_{0},\eta _{1},\eta _{2},\eta _{3})$, these integral operators, acting on
the field $A_{j}(z,\overline{z})$, shall be denoted by $\widehat{L}_{n}(\eta
)$ and $\widehat{J_{n}^{a}}(\eta )$ $\ $
\begin{equation}
\widehat{L}_{n}(\eta )A_{j}(z,\overline{z})=\oint\limits_{C}\left[
\oint\limits_{C_{1}}T(\zeta )(\zeta -z)^{n+1}\delta (\left[ \widetilde{X}%
,\eta \right] +1)A_{j}(z,\overline{z})d\widetilde{X\text{ }}\right] d\zeta
\tag{6.4}
\end{equation}
and the integral operator $\widehat{J_{n}^{a}}(\eta )A_{j}(z,\overline{z})$
will be defined in a completely analogous way. In (6.4) $\widetilde{X}$ is a
point on the ADS hyperboloid with an unit de-Sitter radius and $\delta (%
\left[ \widetilde{X},\eta \right] +1)$ is a \textbf{generalized function},
defined on the set of \textbf{all four-dimensional hyperplanes} (with
coefficients $\eta ^{0}$, $\eta ^{3}$, $\eta ^{1}$, $\eta ^{2}$, determining
their position) and situated on the cone, defined by the \textbf{bi-linear
form }$\left[ \widetilde{X},\eta \right] $
\begin{equation}
-1=\left[ \widetilde{X},\eta \right] =-\widetilde{X}^{0}\eta ^{0}-\widetilde{%
X}^{3}\eta ^{3}+\widetilde{X}^{1}\eta ^{1}+\widetilde{X}^{2}\eta ^{2}\text{
\ .}  \tag{6.5}
\end{equation}
The \textbf{cone} $\left[ \widetilde{X},\eta \right] =0$ is the geometrical
place of \ hyperplanes, separating the hyperplanes, lying \textbf{inside}
the cone, when $\left[ \widetilde{X},\eta \right] <0$ (the present case in
(6.5)) and also the hyperplanes, lying \textbf{outside} the cone, when $%
\left[ \widetilde{X},\eta \right] >0$.

Therefore, it can be obtained that the equality $\left[ \widehat{L}_{n},%
\widehat{J}_{m}^{a}\right] =-m\widehat{J}_{n+m}^{a}$ in terms of the new
''hat'' operators will hold only if
\begin{equation}
\oint\limits_{(\zeta _{1},X_{1})}F(\zeta _{1},\zeta _{2},z)\delta (\left[
X_{1},\eta _{1}\right] +1)d\zeta _{1}dX_{1}=-mJ^{a}(\zeta _{2})(\zeta
_{2}-z)^{m+n}\text{ \ \ \ \ ,}  \tag{6.6}
\end{equation}
and $F(\zeta _{1},\zeta _{2},$ $z)$ is the function $\ $%
\begin{equation}
F(\zeta _{1},\zeta _{2},z)\equiv T(\zeta _{1})J^{a}(\zeta _{2})\left[ (\zeta
_{1}-z)^{n+1}(\zeta _{2}-z)^{m}t_{j}^{m}-(\zeta _{1}-z)^{m}(\zeta
_{2}-z)^{n+1}t_{j}^{n}\right] \text{ \ .}  \tag{6.7}
\end{equation}
\textbf{It is important to note that this is no longer an equality,
identically satisfied as a result of the OPE relations, but represents an
equation}. In other words, the parameters $\eta _{1}^{(0)},\eta
_{1}^{(1)},\eta _{1}^{(2)},\eta _{1}^{(3)}$ (the lower subscript denotes the
parameters, related to first hyperplane integration), ''fixing'' the
hyperplane position, have to be determined in such a way so that the two
sides of (6.6) will be equal.

\bigskip \textbf{Acknowledgement.} The author is greateful to Dr. L.
Alexandrov and Dr. D. M. Mladenov (BLTP, Joint Institute for Nuclear
Research, Dubna) for useful comments and their interest towards this work.
The work on this paper was supported also by a Shoumen University grant
under contract \ 005/2002. The author is greateful also to Dr. P. L.
Bozhilov (Shoumen University, Bulgaria) for the possibility to participate
in this grant.

\end{document}